# Internet sentiment exacerbates intraday overtrading, evidence from Guba.

Peng Yifeng [a,*]

[*] Corresponding author.

[a] School of Management and Economist, Chinese University of Hong Kong (Shenzhen), Shenzhen, Guangdong 518100, China

*E-mail addresses:* yifengpeng@link.cuhk.edu.cn (P. Feng)

## Abstract:

Market fluctuations caused by overtrading are important components of systemic market risk. This study examines the effect of investor sentiment on intraday overtrading activities in the Chinese A-share market. Employing high-frequency sentiment indices inferred from social media posts on the Eastmoney forum "Guba", the research focuses on constituents of the CSI 300 and SZSE 500 indices over a period from 01/01/2018, to 12/30/2022. The empirical analysis indicates that investor sentiment exerts a significantly positive impact on intraday overtrading, with the influence being more pronounced among institutional investors relative to individual traders. Moreover, sentiment-driven overtrading is found to be more prevalent during bull markets as opposed to bear markets. Additionally, the effect of sentiment on overtrading is observed to be more pronounced among individual investors in large-cap stocks compared to small- and mid-cap stocks.

Keywords: Investor sentiment, Chinese stock market, Overtrading, Intraday trading

# 1. Introduction:

In the past decade, the Chinese stock market (also known as the A-share market) has continued to develop amidst pressure and opportunities, with trading volumes constantly increasing. Under the seemingly prosperous surface, the persistently high turnover rate has triggered a series of concerns among scholars about the potential risks of the market (Allen et al., 2024; Setayesh & Daryaei, 2017; Yu et al., 2015). Publicly available information from the Shanghai and Shenzhen Stock Exchanges indicates that over the past decade (CSMAR, 2022), the average turnover rate of the Chinese stock market has been as high as 205.2%, far exceeding the turnover rates of other countries and regions during the same period.

The high trading turnover rate in the stock market generally indicates market activity and high investor participation, reflecting a sufficient amount of liquidity in the market. From another perspective, high trading turnover rates may suggest that market information transmission efficiency is high, and prices can swiftly reflect all available information. However, if a high turnover rate is primarily driven by non-informational trading behaviors, it may have a negative impact on market efficiency. As suggested by Zhang et al. (2019), the high turnover rate in the A-share market is primarily due to the prevalence of overtrading in the market. Most overtrading does not bring expected returns to investors, instead, often causing investors to suffer losses due to irrational behavior. During the market falling process, the herd mentality induced by blind trading behaviors may potentially evolve into panic market crashes (Phan et al., 2018). Therefore, overtrading, which is an element that triggers systemic financial crises, is increasingly drawing attention.

From a perspective of traditional finance theory, the overtrading behavior can be explained as the objective existence of market friction and speculative behavior caused by information asymmetry (Barber & Odean, 2000; Odean, 1999). In an ideal market, friction should not exist, but in reality, transaction costs (such as commission fees and stamp taxes) are present. Excessive trading could increase these costs, which would erode investment returns(Fong, 2014). The information asymmetry hypothesis posits that the existence of information asymmetry can lead to certain investors possessing more information than others, and thus engaging in excessive trading. The underlying reason is that investors do not trust market efficiency and attempt to "beat the market" through short-term trading. With the development of behavioral finance theory, theories such as overconfidence, representativeness bias, and anchoring effects analyze investors' emotions from the perspective of cognitive biases (Kahneman & Tversky, 1972, 2012). Overtrading is a result of a series of irrational decision-making. Globally, cases of market over-fluctuations caused by investors' irrational emotions are not uncommon. On the day of the UK's Brexit referendum, European stock markets fell by about 7%, the largest single-day drop since 2008 (Hassan et al., 2022). In the first month of the Russo-Ukrainian war, investors were concerned about the uncertainty of the future, with the global stock market experiencing significant fluctuations. A large amount of funds flowed into traditional safe haven assets (Priem, 2022). During the COVID-19 pandemic, investors' emotions fluctuated with the recurrence of the virus, and governments' active intervention and control helped to mitigate over-fluctuation (Jiang et al., 2021). These examples have already demonstrated that the analysis of investors' emotions is necessary for understanding over-trading behavior.

Wang et al.(2021) proposed the Securities Investor Information Index to measure investor sentiment in the Chinese stock market, the empirical analysis based on monthly A-share market data from the DAG and VAR models reveals a strong correlation between increases in investor sentiment and an influx of excessive trading behavior. However, Kim et al. (2021) suggest that investor sentiment has a short-term impact. Moreover, the influence of investor sentiment on intraday trading is significantly higher than on overnight trading. High-frequency investor sentiment can predict intraday returns, and high-frequency data performs better than low-frequency data (Seok et al., 2021; Sun et al., 2016). This study will leverage data from the guba.eastmoney.com to construct a high-frequency investor sentiment index and specifically analyze the impact of investor sentiment on overtrading behavior in the A-share market.

The potential contributions and innovations of this paper are as follows:

1. By using web crawlers to obtain posting text data in the forum of Eastmoney forum “Guba”, we construct a high-frequency social media investor sentiment indicator based on text analysis methods, which covers an hour-long span. The A-share market is abundant in retail investors, and online media is an important platform for individual investors to express emotions in the stock market. Eastmoney forum “Guba” is the largest and most influential stock forum for retail investors in China, and extracting financial text data from the forum can more accurately estimate the high-frequency changes in investor sentiment and, based on this, predict the excessive trading behavior of individual investors.
2. This paper differentiates between the effects of investor high-frequency sentiment on the excessive trading behavior of institutional investors and individual investors. Personal retail investors may be more susceptible to irrational sentiments than institutional investors. The prevailing short-term trading in the A-share market magnifies the impact of emotions-induced excessive trading behavior. This paper analyzes the causes of excessive trading behavior among different types of investors, enriching the research on the behaviors of A-share market investors. At the same time, it suggests measures for investors to prevent and overcome irrational excessive trading behavior, as well as providing insights for regulatory institutions in managing and supervising the market.

3. This paper distinguishes the moderating effects of investor high-frequency sentiment on excessive trading behavior among investors based on different market environments and different market capitalizations. From an empirical perspective, this paper confirms the asymmetrical role of high-frequency investor sentiments in predicting short-term excessive trading behavior in the A-share market. This provides a reference for the subsequent investment activities in the A-share market.

# 2. Literature Review and Hypothesis Development

## 2.1 The Impact of Investor Sentiment on Overtrading Behavior

Overtrading is widely prevalent in major financial markets around the world, as demonstrated by Fong's empirical research (2014), which suggests that overtrading in the US stock market does not necessarily lead to high returns in investors' expectations. Liu et al. (2022) collected thousands of survey questionnaires targeting individual investors in the A-share market, and combined the results with actual investment behavior, the study found that the average monthly turnover rate among A-share individual investors is 84.8%, while the post-fee monthly net yield rate is -0.2%. Findings suggest that frequent trading among these investors has yielded sub-optimal returns.

Chakraborty (2003) pointed out that the overtrading brought about by over-the-counter trading is the primary culprit responsible for return erosion. High-frequency trading not only increases transaction costs, but also increases the likelihood of subsequent irrational trading. Currently, there is a general consensus that the significant psychological factors leading to investor over-trading are investors' "overconfidence" (Bregu, 2020). Hoffmann et al. (2016) believe that investors' overconfidence relies on intuitive judgment as well as they are more likely to traverse across various high-risk financial products. They tend to overly trust their market judgment and information-processing capability, while attribute investment failures to external objective factors. This cognitive bias of extrapolating personal returns excessively to the future will further provide more trading reasons for confident investors, leading to a positive feedback loop. Good returns often make investors more confident, leading to more active trading in subsequent transactions. However, when faced with setbacks, overly confident investors seem to struggle to convince themselves to stay in. Gottesman et al. (2023; 2022) view overconfidence as a disadvantage in combating depressed and declining markets. When the market enters a slump, even with signs of a stock market crash, investors with an overconfidence investment style are more likely to quickly sell off and withdraw from the market. This stress often misses out on subsequent market recovery.

The relationship between investor sentiment and excessive trading is evident. Studies conducted by Parveen et al. (2020; 2021) found a negative correlation between investor sentiment and future stock returns, and this influence exhibits heterogeneity. Investor sentiment has a more direct impact on emerging markets, such as India and Pakistan in South Asia, while its effect on developed markets, such as the United States and the Eurozone, is more persistent. Smales (2017) presents a different explanation for overtrading behavior, the implied volatility index (VIX) remains a very effective tool to assess investor sentiment indices in specific markets nowadays. The underlying principle is the accumulation of panic caused by investors' increase in uncertainty regarding future events, further leading to the proliferation of excessive trading behavior. Studies from mainstream markets indicate that during periods of economic downturn or when market sentiment is at its lowest (Wang et al., 2021), the impact of sentiment on market returns is greater, especially for stocks most susceptible to speculative demand.

Based on the aforementioned research, this paper proposes three areas to explore regarding the impact of investor sentiment on excessive trading behavior: market returns, investors' overconfidence sentiment, and market trading volume.

Market Return: Empirical evidence from global mainstream markets indicates a negative correlation between market sentiment indices and future returns (Wang et al., 2021). Chung et al. (2012) further explored the heterogeneity of investor sentiment and stock returns under different market states. In the expansion phase, the predictive ability of the sentiment index for returns is stronger. Wang(2024) further investigates the negative correlation between sentiment and returns during both the non-trading and trading periods. The impact of sentiment on intraday returns is significantly greater than overnight returns. On the other hand, the changes in stock returns can also influence investors' risk preferences (Fong, 2013; Z. He, 2022). This further affects future investor decisions.

Overconfidence: When "overconfident" investors gain market returns from their decisions, they tend to attribute the investment returns to their judgment, and view this "surpassing the market" judgment as a reflection of their abilities (Bregu, 2020). Based on the observations of investors' mindset and actual investment behavior by Merkle et al. (2020; 2016; 2022), it has been observed that investors who exhibit overconfidence tend to engage in stock trading activities more frequently than those who exhibit long-termism and rationality. Asaad(2020) also corrected the view that overconfident individuals blindly ignore risks and make judgments and actions due to their overconfidence, suggesting that overconfident investors may inadvertently expose themselves to risks, rather than recklessly disregarding them.

Volume: Liu(2015) posits that there is a general positive correlation between investor sentiment and market liquidity, and the increase in market liquidity can also lead to an increase in market volume. Lee et al. (2002) further elucidated the reasons behind the increase in market liquidity caused by investor sentiment. When investor sentiment is high, it contributes to market noise, and at the same time, it exhibits a positive correlation with changes in excess returns. This further validates the argument made by Black(1986) about the relationship between market noise and market profitability opportunities. Willman et al.(2006) further explained the impact of market noise on volume from the perspective of traders. When noise increases, trading is not simply conducted for profit, but also generates information, accelerates learning, creates commitment, and enhances social capital, all of which allow traders to maintain their long-term survival in the market. When market profitability opportunities increase, both overly confident investors and partially rational investors will try to exploit these opportunities to attempt to generate excess profits. These pursuit of profits naturally lead to increases in trading volume and the occurrence of overtrading behaviors(Bloomfield et al., 2009; Verma & Verma, 2007).

Based on the above studies, market participants, characterized as limited rational agents, are often swayed by emotions, leading to an overestimation of the accuracy and rationality of their judgments and decisions, resulting in excessive trading. Based on the aforementioned analysis, this paper proposes the following hypotheses:

Hypothesis 1 (H1): The fluctuations in investor sentiment are positively correlated with the excessive trading activity levels observed in the Chinese A-share market.

## 2.2 Investor sentiment influences the heterogeneity of overtrading of institutional investors and individual investors

compared to value investment and long-termism. Short-term high-frequency trading can more accurately reflect the dispositions of institutional investors and retail investors in the A-share market (Ng & Wu, 2006; Leightner, 2011). Numerous studies have confirmed that short-term trading, or high-frequency trading, is prone to excessive trading behaviors. The reasons can be explained from various perspectives, such as the increased transaction costs caused by frequent trading, the surge in excessive trading due to noise trading, and the reduction in rational thinking processes driven by emotional factors (Fong, 2014; Ke & Zhang, 2020; Barber & Odean, 2000; Bloomfield et al., 2009). Institutional investors stand out in terms of substantial capital, high-level stock selection abilities, and stronger professionalism. Moreover, large institutions can often obtain internal trading information or insider news more quickly, which grants them the ability to react to market information and investor sentiment changes ahead of time (Bricker & Markarian, 2015). In contrast, individual investors exhibit low professionalism, small capital, susceptibility to emotional contagion, and a lack of investment experience. These characteristics determine that they would make different assessments of the same investor sentiment and consequently produce different trading behaviors.

On one hand, institutional investors typically believe that information disclosure is more thorough while institutions often perceive themselves to have stronger informational advantages and professional capabilities compared to individual investors. Moreover, institutional investors can convert these advantages into excess returns. This mindset lays the groundwork for overconfidence. To pursue market opportunities, institutions frequently adjust their portfolios, which under the influence of investment sentiment, may exhibit more severe overtrading than individual investors (Huang et al., 2024; Fong, 2014; Phan et al., 2018). On the other hand, individual investors may also be prone to overconfidence due to their relatively limited investment experience. This mindset can lead to aggressive buying or panic selling under extreme circumstances (Gottesman & Morey, 2023). Simultaneously, individual investors are more susceptible to external environment and emotional influence. Their attention and rational boundaries are narrower, making it more challenging to acquire and process information accurately. This can lead to the formation of information cocoons, and they are more prone to overtrading under the influence of emotions.

Based on this analysis, this paper presents two opposing hypotheses:

Hypothesis 2a: Compared to individual investors, investor sentiment has a greater impact on the overtrading behavior of institutional investors.

Hypothesis 2b: Compared to institutional investors, investor sentiment has a stronger impact on the overtrading behavior of individual investors.

# 3. Design of Research

## 3.1 Sample Selection and Data Sources

The CSI 300 Index covers the 300 largest and most liquid stocks in the Shanghai and Shenzhen stock markets, representing a significant segment of the Chinese equity market. The CSI 500 Index, on the other hand, encompasses the 500 stocks from Shanghai and Shenzhen markets that are not included in the Hang Seng China Enterprises Index, representing the overall condition of medium-sized stocks in the Chinese equity market. Together, these indices cover the 800 largest stocks, accounting for a large portion of the total market value in Shanghai and Shenzhen stock markets, thus they possess strong market representativeness.

This paper takes the constituent stocks of the Hang Seng China Enterprises Index and the CSI 500 Index as research subjects. Through web scraping, we collected posts from the Oriental Fortune Securities Forum on the 1st of January, 2018, and the 30th of December, 2022. We then obtained the intraday trading data of each stock from the Wind database by comparing the timestamps of the posts to the time period of the A-share market trading hours. The sampling period spans 1215 trading days, and the frequency of text processing was set to one hour, with each trading day divided into four segments: 9:30-10:30, 10:30-11:30, 13:00-14:00, and 14:00-15:00. A total of 4860 sample segments were obtained.

In order to ensure the credibility of the empirical results, the data cleaning process will be carried out in accordance with the following rules after the use of the network crawler to capture meta-data: 1. To avoid the impact of few forum discussions on the quality of data caused by investors in individual stocks paying too much attention, samples are removed for stocks whose interval with no discussions for an hour or more accounts for more than 10% of the total interval; 2. To avoid the impact of long-term suspension stocks on the market trend caused by the price fluctuations after resumption, stocks with a cumulative suspension period exceeding 30 days are removed from the sample; 3. To avoid the correlation effect between the index and individual stocks (also known as the "index effect" (Y. He & Wang, 2015)), samples are adjusted where more than two records are recorded within the sampling period, thereby removing samples; 4. The number of robot posts and meaningless posts in Chinese Internet financial forums has increased year by year (Li et al., 2019), and since the upgrade of the Ba Ba forum in 21, the official has deployed an automatic post summary consulting robot, which this paper will filter out specific user IDs (such as "Ask-Sectary Robot", "AI Summary") to avoid repeated data collection.

Furthermore, this paper also deletes posts with abnormal posting times, posting times missing, announcements and inquiries that cannot be replied. Ultimately, 325 stocks with 4,924,614 forum posts were obtained.

## 3.2 Analysis of Model Selection and Setting

Based on the actual intraday trading time, to analyze the influence of high-frequency investor sentiment on the excessive trading behavior of intraday investors, this paper establishes the following base model:

$$ET_{i,h} = \alpha + \beta S_{i,h-1} + \varepsilon_{i,h} \quad h = 2,3,4 \tag{1}$$

In Formula (1), $ET_{i,h}$ represents the excess turnover rate of stock $i$ during the trading day at Hour $h$, and $S_{i,h-1}$ represents the investor sentiment measure for stock $i$ during the time period from Hour $h-1$ to Hour $h$ of the trading day. The focus of this study is on the interplay between investor sentiment and excessive trading within a trading day, hence the data for the regression model will be collected from hour 2 of each trading day onwards.

### 3.3 Details of model variables

#### 3.3.1 Explanatory variable: investor sentiment index

##### 3.3.1.1 Dictionary corpus construction

In this study, we construct a specialized dictionary corpus by utilizing multiple open-source Chinese sentiment lexicons, such as the Positive and Negative Word Dictionary compiled by Tsinghua University's Professor Li Jun, the Hownet Chinese sentiment dictionary, the Chinese sentiment lexicon compiled by Dalian University of Technology, and the NTUSD Simplified Chinese Emotion Dictionary from Taiwan University, as the foundational emotional lexicon corpus. Moreover, due to the data being sourced from social media platforms, we also utilize the BosonNLP Chinese sentiment dictionary constructed based on data from Weibo, news, forums, etc. Additionally, we utilize the EmoDict dictionary based on internet slang as a benchmark in our sentiment analysis.

We process the textual data from posts using JIEBA's precise mode for segmentation and word frequency count. After obtaining the word frequency count results, we manually label and add the emotional words with a frequency greater than or equal to 80 in the corpus. Finally, the constructed corpus of dictionary data includes 50,182 negative words and 39,767 positive words.

3.3.1.2 Emotional Scoring

After the text data is segmented, the weight of the sentiment words is determined according to the word dictionary method. Positive sentiment words are weighted with 1, while negative sentiment words are weighted with -1, and neutral words are weighted with 0. In the context of Chinese, double negation often implies positive sentiment. This can be modeled as the number of negative words will influence the semantic tendency. This research will focus on the number of negative words after segmentation and construct the following formula for text emotion calculation:

$$S_{i,h,j} = \frac{\sum_{k=1}^{E} W_k}{E}(-1)^n, n \in N, E \in N_+ \quad (2)$$

$S_{i,h,j}$ in formula (2) represents the sentiment value of the stock $i$ 's post in the period $h$. $W_k$ is the weight of the kth sentiment word, and $E$ is the total number of negative sentiment words in the post. $N$ represents the set of natural numbers, $N_+$ represents the set of positive integers. For most contexts, $n \leq 2$ holds.

3.3.1.3 Investor Sentiment Measurement

Based on the calculation of the sentiment value of a single post, the investor sentiment over the entire period can be calculated by obtaining the average sentiment value of all posts within each period, as shown in formula (3):

$$S_{i,h} = \frac{\sum_{j=1}^{N} S_{i,h,j}}{N} \quad (3)$$

In formula (3), $N$ represents the total number of posts for stock $i$ in period $h$, while $S$ represents the investor sentiment measurement for the stock $i$ in period $h$. To verify the accuracy of the word dictionary approach in calculating text emotion, approximately 5,000 posts were randomly selected and their emotional tendencies were assessed manually. The method for assigning emotion scores is the same as in the "Emotion score calculation" section. After comparing the text emotion calculated by the word dictionary approach with that obtained manually, it was found that the accuracy rate of the word dictionary approach is approximately 80.21%.

### 3.3.2 Explained variable: excessive trading behavior of investors

Inspired by the use of excess transaction volume as a quantitative indicator for excessive trading behavior in the articles of Zhang et al. (2014; 2019), this paper calculates the current turnover ratio by dividing the transaction volume by the transaction price and the corresponding total shares of the stock at the current period, and then calculates the excess turnover ratio by subtracting the current turnover ratio from the historical turnover ratio.

$$T_{i,h} = \frac{\sum_{j=1}^{h} V_{i,j} * P_{i,j}}{C_{i,h}} \tag{4}$$

$$ET_{i,h} = \frac{T_{i,h} - \frac{\sum_{j=1}^{M} T_{i,h-j}}{M}}{\frac{\sum_{j=1}^{M} T_{i,h-j}}{M}} \tag{5}$$

In Formula (4), $V_{i,j}$ represents the transaction volume of stock $i$ during period $j$, $P_{i,j}$ represents the transaction price of stock $i$ during period $j$,, and $C_{i,h}$ represents the total number of shares of stock $i$ during period $h$. In Formula (5), $T_{i,h}$ represents the turnover ratio of stock $i$ during period $h$, and $ET_{i,h}$ represents the excess turnover ratio of stock $i$ during period $h$.

As mentioned in the literature review, excessive trading behavior can be divided into excessive trading behaviors of institutional investors and excessive trading behaviors of individual investors. This paper classifies each transaction into a large order, a medium-sized order, a small order, and an ultra-small order based on the transaction amount and the number of shares traded, respectively. When the transaction is classified into a large order and a medium-sized order (that is, the total amount of the transaction exceeds 200,000 yuan or the number of shares traded exceeds 100,000), it is deemed to be an institutional investor's trading behavior. Conversely, it is deemed to be an individual investor's trading behavior.

3.4 Descriptive statistical analysis

As shown in Table 1, although investor sentiment in the sampling period did experience extremely optimistic conditions (with a maximum value of 13.6784), the overall average expectation in the stock market during the sampling period was more pessimistic (with a mean value of -0.2904). The excessive trading behavior of investors is somewhat scattered in time, but the maximum value of the excess trading ratio reached 77.6541, indicating extreme excessive trading behavior in the data at a certain moment. When looking at investors separately, the excess trading ratios of institutional investors are higher than those of individual investors across all metrics, indicating that the excessive trading behavior of institutional investors is more frequent and more intense than that of individual investors.

Table 1 Descriptive statistics of explanatory variables and explained variables

| | Average | Std | Maximum | Minimum |
|---|---|---|---|---|
| Sentiment | -0.2904 | 0.3122 | 13.6784 | -6.8785 |
| ET(Institute) | 0.0803 | 1.4021 | 132.8870 | -1.0000 |
| ET(Retail) | 0.0377 | 0.8232 | 74.9623 | -1.0000 |

This table shows the descriptive statistics of the explanatory variable (investor sentiment) and the explained variable (excessive trading behavior). According to the formula, the minimum excess turnover rate of investors is -1, while the actual sentiment minimum is -4.736

# 4. Empirical results and analysis

4.1 Investor sentiment analysis of the intraday effects of overtrading behavior in institutional investors and individual investors

The regression results for excessive trading behavior across all samples are shown in Table 2. For the first, second, and third hour, the investor sentiment in the previous hour has a significant effect on the future hour’s excess turnover rate at the 1% level. This implies that the investor sentiment of the previous hour can stimulate or facilitate excessive trading behavior in the next hour. This result can be interpreted as investors being influenced by the emotions expressed by other investors on online platforms, which ultimately leads to excessive trading behavior. This effect is particularly noticeable in the short term. Investors may overreact to new information on online social networks, leading to overvaluing or undervaluing stocks, resulting in irrational trading behaviors. This finding validates Hypothesis 1, and in terms of the three time periods, the excessive trading behavior in the second hour of the trading day is most influenced by the investor sentiment in the previous hour. This reflects that excessive trading behavior is more likely to occur in the morning sample period than in the afternoon, which may be related to the unique T+1 trading mechanism in the A-share market. According to empirical studies conducted by Guo et al. (2017; 2012), A-shares typically exhibit a pattern of lower opening prices in the morning and higher closing prices in the afternoon. From a profit-seeking perspective, the morning is a better time to initiate a position than the afternoon.

A further observation of the data reveals that the investor sentiment from 1st, 2nd, and 3rd hours is significantly associated with the over-trading rate of institutional and individual investors over the next hour, at a level of 1%. The positive effect of investor sentiment on all investors' excessive trading behavior is further validated by Hypothesis H1. Additionally, it can be observed that the impact of investor sentiment on the excessive trading behavior of institutional investors is greater than that of individual investors, and the magnitude of this impact is approximately 1.8 times that of individual investors. The cause behind this phenomenon may be the significant information asymmetry in the A-share market, where institutional investors can obtain more information and channels, and possess insider information. In a short period, institutional investors with information advantages can better grasp market trends compared to individual investors with information disadvantages. Moreover, the team of professional investors and their brainstorming can also be displayed as advantages. Therefore, institutional investors are more susceptible to social media investor sentiment and are likely to engage in excessive trading behavior. This confirms Hypothesis H2a.

Table 2 Regression results of full sample investor sentiment on investors' overtrading

| | 10:30-11:30 ET | 13:00-14:00 ET | 14:00-15:00 ET |
|---|---|---|---|
| Last hour sentiment | 0.1305*** | 0.0911*** | 0.0673*** |
| Current sentiment | (54.23) | (16.22) | (13.37) |
| $\alpha$ | -0.1996*** | -0.1008*** | -0.0741*** |
| | (-103.24) | (-67.72) | (-44.31) |
| $N$ | 200,000 | 200,000 | 200,000 |
| Adj R-squared | 0.0816 | 0.0553 | 0.0489 |
| F | 3.381 | 5.282 | 8.013 |
| Wald chi$^2$(1) | 745.32 | 529.11 | 120.21 |

Note: *, **, and *** are significant at the 10%, 5%, and 1% levels, respectively, and the values in brackets are T-values.

Table 3 Regression results of full sample investor sentiment on investors' overtrading

| | Institute investor | | | Retail investor | | |
|---|---|---|---|---|---|---|
| | 10:30-11:30 ET | 13:00-14:00 ET | 14:00-15:00 ET | 10:30-11:30 ET | 13:00-14:00 ET | 14:00-15:00 ET |
| Last hour | 0.3745*** | 0.1336*** | 0.1303*** | 0.2441*** | 0.0972*** | 0.0439*** |
| Current | (30.01) | (13.87) | (10.01) | (12.51) | (8.22) | (7.67) |
| $\alpha$ | 0.5039*** | -0.0808*** | -0.1241*** | 0.3861*** | -0.1173*** | -0.0923*** |
| | (64.32) | (-13.54) | (-22.11) | (50.92) | (-61.33) | (-66.71) |
| $N$ | 200000 | 200000 | 200000 | 200000 | 200000 | 200000 |
| Adj $R^2$ | 0.0308 | 0.0194 | 0.0107 | 0.0291 | 0.0097 | 0.0089 |
| F | 2.00 | 2.55 | 1.90 | 7.22 | 2.31 | 5.95 |
| Waldchi$^2$ | 562.03 | 109.11 | 122.21 | 223.16 | 107.99 | 108.31 |

Note: *, **, and *** are significant at the 10%, 5%, and 1% levels, respectively, and the values in brackets are T-values.

### 4.2 Investor Sentiment Analysis of Overtrading Behaviors under Different Market Conditions

The stock market can be divided into two states, bull and bear, based on the NBER (National Bureau of Economic Research) business cycle measurement method proposed by Pagan and Sossounov (2003). According to the NBER method, a one-way trading cycle of 5 months is set for the Shanghai Composite Index and Shenzhen Component Index. The peak and trough of the cycle are considered the turning points of the transition between bull and bear market conditions. This study uses model (1) to analyze the excessive trading behaviors based on samples from bull and bear markets, and the results are shown in Table 4.

Table 4 Regression results of investor sentiment under bull/bear markets on overtrading behavior

| | Sentiment last hour | ET in 2nd hour | ET in 3rd hour | ET in 4th hour |
|---|---|---|---|---|
| Panel for full sample | Bull Market | 0.2107*** (44.18) | 0.0923*** (26.31) | 0.0587*** (9.22) |
| | Bear Market | 0.0593*** (14.22) | 0.0364*** (12.03) | 0.0149*** (5.32) |
| Panel for institute | Bull Market | 0.3990*** (15.70) | 0.1603*** (10.19) | 0.1258*** (9.73) |
| | Bear Market | 0.2210*** (8.90) | 0.0764*** (3.37) | 0.0449*** (3.87) |
| Panel for retail | Bull Market | 0.2532*** (18.21) | 0.0907*** (12.34) | 0.0787*** (11.55) |
| | Bear Market | 0.1532*** (6.63) | 0.0235*** (3.35) | 0.0349*** (4.24) |

It can be observed that both the results of the bull and bear market support hypotheses H1 and H2a. The excessive trading behavior induced by investor sentiment in the bull market is significantly greater than in the bear market. From the perspective of transaction volume, the high investor sentiment in the bull market attracts more investors to open accounts and generates higher transaction volume and more irrational investments. However, in the bear market, most investors are "stuck" with their holdings, and they have already recognized that the stock prices are overvalued and the possibility of profit is small. Hence, their trading activeness and the quantity of their remaining funds are lower than in the bull market, resulting in a lower level of excessive trading. From a psychological standpoint, the panic in the bear market will prompt investors to consider whether to conduct a sell-off and limit loss. Such actions reduce account funds and increase investment costs, and due to the effect of the disposition effect, investors still hold hopes for the market and are reluctant to immediately sell stocks that have incurred losses. This causes hesitation in their investment decisions, thereby reducing their investment behavior in the bear market compared to the bull market.

### 4.3 Analysis of the intraday effects of overtrading behavior in different-sized stocks based on investor sentiment

The sample stock was divided according to the size of its circulating shares. Specifically, stocks with a circulating shares exceeding 100 billion were classified as large-cap stocks, while stocks with a circulating shares between 10 and 100 billion were considered mid-cap stocks, and stocks with a circulating shares less than 10 billion were classified as small-cap stocks. The data of the large, medium, and small-cap stocks were used to regress the model (1):

Table 5 Regression results of investor sentiment on overtrading under different market value stocks

| | Sentiment last hour | ET in 2nd hour | ET in 3rd hour | ET in 4th hour |
|---|---|---|---|---|
| Panel for full sample | Large-cap stock | 0.1297*** (13.18) | 0.0823*** (9.86) | 0.0687*** (6.22) |
| | Medium-cap stock | 0.1357*** (23.04) | 0.1007*** (12.03) | 0.0749*** (5.32) |
| | Small-cap stock | 0.1187*** (8.80) | 0.0756*** (6.35) | 0.0880*** (5.32) |
| Panel for institute | Large-cap stock | 0.4385*** (17.31) | 0.1415*** (12.87) | 0.1458*** (8.42) |
| | Medium-cap stock | 0.3315*** (18.29) | 0.1009*** (13.02) | 0.1630*** (4.16) |
| | Small-cap stock | 0.3715*** (8.29) | 0.2164*** (2.65) | 0.1490*** (2.16) |
| Panel for retail | Large-cap stock | 0.2748*** (17.30) | 0.1039*** (13.87) | 0.0400*** (5.02) |
| | Medium-cap stock | 0.2001*** (12.25) | 0.1055*** (9.65) | 0.1790*** (12.07) |
| | Small-cap stock | $0.1532^{***}$<br>(9.36) | $0.0875^{***}$<br>(5.09) | $0.0477^{***}$<br>(6.68) |

The regression results in Table 5 indicate that both the hypothesis H1 and H2a are supported by the results of all three types of stocks, large stocks, medium stocks, and small stocks. Comparison reveals that investor sentiment has a greater impact on excessive trading behavior in large stocks than in medium and small stocks. This suggests that most investors prefer to trade stocks belonging to large, risk-averse, and promising companies. One possible inference is that due to the widespread impact of the Sino-US trade frictions in 2018 and the COVID-19 pandemic in 2020, most investors have reduced their risk appetite and are increasingly willing to trade stocks of large companies with low risk and promising prospects. Conversely, in the third hour and fourth hour of trading days, where individual investors and institutional investors are involved, institutional investors in small stocks exhibit more severe excessive trading behaviors than those in large stocks. This may be due to the smaller amount of funds in small stocks and hence their greater susceptibility to "speculation" from institutional investors and market makers. In the same investor sentiment, stocks of small companies often exhibit higher volatility and operability, providing institutional investors with higher potential returns. The high volatility and operability, in turn, can lead to higher trading volume and thereby more noticeable excessive trading behaviors compared to individual investors. Hence, institutional investors in small stocks are more sensitive to investor sentiment.

The results of the above three robustness tests are not significantly different from the regression results in empirical studies. The model has strong explanatory power, which can support all the research hypotheses. This suggests that the original model has passed robustness tests.

## 5. Robustness test

This paper employs three methods for robustness testing: lagged multiple periods regression, increased control variables, and a change in the investor type classification method. Regarding the lag, the investor sentiment index in Model (1) was adjusted to lag two and three periods respectively, considering that the impact of investor sentiment on investor overtrading behavior is transmitted over time. A White test and LM test were performed for the lag regression to minimize the possibility of model endogeneity. In terms of control variables, the company PB, Market Risk Premium Factor, and Market Return Ratio (MRE) were added to Model (1). The threshold for determining whether total transaction amount is greater than 200,000 RMB was changed to whether total transaction amount is greater than 500,000 RMB, when segmenting overtrading behavior of institutional and individual investors.

The results of the three robustness tests are consistent with the regression results in the empirical study, indicating a strong explanatory power for the model and support for all research hypotheses. This demonstrates that the original model has passed robustness tests.

## 6. Conclusion and limitation

This study is based on the posts and stock trading data released on the Oriental Fortune Stock Bar Forum from January 1, 2018 to December 30, 2022, and utilizes hourly intervals within each trading day as the study frequency. This study aims to establish a metric for measuring investors' high-frequency emotions and an indicator for overtrading behavior. The paper analyzes the influence of investors' emotions on different types of investors, different market trends, and the market capitalization of various companies on overtrading behavior. The following conclusions are drawn

from the study:

Investors' emotions significantly impact investors' intra-day overtrading behaviors, with the influence on institutional investors stronger than that on individual investors. In the bull market, the intra-day overtrading behaviors caused by investors' emotions are more pronounced than in the bear market. The impact of investors' emotions on individual investors' intra-day overtrading behaviors in large-cap stocks is greater than in small-cap stocks. However, institutional investors exhibit a more severe form of overtrading behavior due to investors' emotions in small-cap stocks.

When quantifying emotion scores, this paper simply polarizes emotions. If advanced neural networks and deep learning techniques are used, emotions can be further converted into continuous numerical scores. The Bi-LSTM model, which is more widely used in the industry at present, can be used to process the text more carefully and then accumulate the emotion score(Bhuvaneshwari et al., 2022; Minaee et al., 2019).

At the same time, future studies can further explore the mechanism of investor sentiment's influence on excessive trading behavior, such as the different effects of different types of investor sentiment (such as optimism, pessimism, greed, panic, etc.) on excessive trading behavior.